\documentclass[prl,twocolumn,showpacs,amsmath,amssymb,superscriptaddress]{revtex4-2}
\usepackage{graphicx}
\usepackage{hyperref}
\usepackage{xcolor}
\usepackage{color}
\usepackage[all,cmtip]{xy}
\usepackage{amsmath,amsthm} 
\usepackage{amsfonts}
\usepackage{relsize}
\makeatother

\begin{document}

\title{Chiralities of nodal points along high symmetry lines with screw rotation symmetry}
\author{Rafael Gonz\'{a}lez-Hern\'{a}ndez}
\email{rhernandezj@uninorte.edu.co}
\affiliation{Departamento de F\'{i}sica y Geociencias, Universidad del Norte, Km. 5 V\'{i}a Antigua Puerto Colombia, Barranquilla 080020, Colombia}
\affiliation{Institut f\"ur Physik, Johannes Gutenberg Universit\"at Mainz, D-55099 Mainz, Germany}
\author{Erick Tuiran}
\email{etuiran@uninorte.edu.co}
\affiliation{Departamento de F\'{i}sica y Geociencias, Universidad del Norte, Km. 5 V\'{i}a Antigua Puerto Colombia, Barranquilla 080020, Colombia}
\author{Bernardo Uribe}
\email{bjongbloed@uninorte.edu.co}
\email{uribe@mpim-bonn.mpg.de}
\affiliation{Departamento de Matem\'{a}ticas y Estad\'{i}stica, Universidad del Norte, Km. 5 V\'{i}a Antigua Puerto Colombia, Barranquilla 080020, Colombia}
\affiliation{Max Planck Institut f\"ur Mathematik, Vivatsgasse 7, 53115 Bonn, Germany}

\date{\today}

\begin{abstract}

Screw rotations in nonsymmorphic space group symmetries induce the presence of hourglass and accordion shape band structures along 
screw invariant lines whenever spin-orbit coupling is non-negligible. These structures induce topological enforced Weyl points on the band intersections. In this work we show  that the chirality
of each Weyl point is related to the representations of the cyclic group on the bands that form the intersection. To achieve this,
we calculate the Picard group of isomorphism classes of complex line bundles over the 2-dimensional sphere with cyclic group action,
and we show how the chirality (Chern number) relates to the eigenvalues of the rotation action on the rotation invariant points. Then
we write an explicit Hamiltonian endowed with a cyclic action whose eigenfunctions restricted to a sphere realize the equivariant line bundles described before. 
As a consequence of this relation, we determine the chiralities of the nodal points appearing on the hourglass and accordion shape structures on screw invariant lines
of the nonsymmorphic materials PI$_3$ (SG: P6$_3$), Pd$_3$N (SG: P6$_3$22), 
 AgF$_3$ (SG: P6$_1$22) and AuF$_3$ (SG: P6$_1$22),
 and we corroborate these results with the Berry curvature and symmetry eigenvalues calculations for the electronic wavefunction.
\end{abstract}
\maketitle

%\twocolumngrid
\section{Introduction}

An interesting research topic in topological condensed matter is the study of energy band crossings, particularly those protected by crystal symmetries \cite{Colloquium-topological-insulators,RevModPhys.88.021004}. The presence of energy degeneracies near the Fermi level can lead to particular topological phases known as Dirac and Weyl semimetals \cite{Weyl-and-dirac-semimetals}.  In Dirac semimetals the energy band crossings are four-fold degenerate at high symmetry points and they are protected by certain symmetries.  These materials can be viewed as 3D analogues of graphene, with Dirac cones in the Brillouin zone as it has been confirmed experimentally \cite{Discovery-Dirac-Semimetal, Experimental-Realization-Dirac,Xiong413}. However, if inversion symmetry or time-reversal symmetry (or possible both) is broken, the Dirac point splits to form a pair of two-fold band degeneracies, which can be located at the Fermi level \cite{Topological-Semimetals,Topological-Materials}. These kinds of materials are known as Weyl semimetals and magnetic Weyl semimetals, depending of the specific broken symmetry \cite{Discovery-of-Weyl-semimetals,Discovery-Weyl-fermion,MWSM,Wang2018}. Independent of the system, an important property of these band degeneracies (or Weyl points) is that they have a quantized monopole charge or chiral charge living in the reciprocal space. The topological charge (chirality) usually is found to be $\pm$1 due to the linearity of the Hamiltonian in each direction of the three-dimensional reciprocal space \cite{PhysRevX.5.011029}. These Weyl points may induce an exotic family of phenomena such as Fermi arcs surface states \cite{Discovery-Weyl-fermion,PhysRevB.83.205101}, giant anomalous and spin hall effects \cite{recent-dev-transport,PhysRevResearch.2.013286,PhysRevLett.117.146403,AHE-in-WS}, chiral anomalies \cite{Observation-Chiral-Anomaly-Induced,chiral-anomaly,Zhang20162017}, and giant responses to external stimulus \cite{tranport-in-wsm,Quantized-circular}. 

In the case of non-magnetic Weyl semimetals, the presence of nonsymmorphic symmetries, such as screw rotations or glide reflections, can generate band crossings in high-symmetry lines or planes of the Brillouin zone \cite{Symmetry-demanded-topological,Nonsymmorphic-symmetry-required-band-crossings,semimetals-in-nonsymmorphic-lattices}. In particular, Weyl points in hourglass and accordion shape energy dispersions along  invariant lines of the Brillouin zone are protected by the combination of screw rotation symmetry and time-reversal symmetry, and furthermore they are topologically stable \cite{Topological-crossings-trigonal,Topological-crossings-hexagonal,Gonzalez-Tuiran-Uribe,n-hourglass}. Recently, it has been observed that these kinds of symmetries can also produce exotic fermions with 3-fold, 4-fold and 6-fold degeneracies at the time reversal invariant points of the Brillouin zone \cite{Beyond-Dirac-and-Weyl-fermions}. The energy bands of these multifold fermions have higher topological charge $\chi$=$\pm$2, $\pm$4, $\pm$6 with more Fermi arcs and hence enhance many of the phenomena associated with $\chi$=$\pm$1 Weyl fermions  \cite{Beyond-Dirac-and-Weyl-fermions,Topological-chiral,LV2017,PhysRevLett.108.266802,Twofold-quadruple-Weyl,maximal-Chern-numbers}. In addition, double and triple composite Weyl points have been also predicted in the invariant lines of the Brillouin zone for nonsymmorphic materials \cite{double-Weyl,Vanderbilt}.

On the other hand, the calculation of Weyl point chirality is involved and requires the computation of the Berry curvature flux over a sphere in reciprocal space, with a dense $k$-mesh integration \cite{Berry-phase}. This integration is usually carried out in the Wannier representation, which can unintentionally break off wavefunction symmetries, as it has been noted in \cite{MLWF,Symmetry-adapted-Wannier}.  Therefore, motivated by these facts, we have demonstrated that it is possible to use directly the eigenvalues of the symmetry operators to find the chirality of the Weyl points in nonsymmorphic materials. We propose a Hamiltonian model which produces higher Weyl chirality in agreement with the eigenvalues of symmetry operators and Berry curvature flux calculations. We also provide mathematical 
proof that shows that this Hamiltonian exhausts all the possibilities for the topological structures that may appear at a Weyl point endowed with rotational symmetry.  Finally, we corroborate the formation of triple/double Weyl points with cubic/quadratic dispersion and higher chiralities $\pm$3/$\pm$2 in hexagonal nonsymmorphic materials such as PI$_3$ (SG: P6$_3$), Pd$_3$N (SG: P6$_3$22), AgF$_3$ (SG: P6$_1$22), AuF$_3$ (SG: P6$_1$22).  

We finish this introduction by noting that the idea of using the eigenvalues of the rotation
operator on the energy bands to detect the chiralities of the nodal points have been explored before.
In particular, in \cite{Multi-Weyl-Topological} a table with compatibility conditions for the chiralities
is presented after studying the behaviour of the first terms of the Hamiltonian with respect to
representations of the bands. Our approach instead uses equivariant K-theory to determine all equivariant Hamiltonians up to adiabatic deformation, which in turn allows us the determine an explicit formula relating the 
congruence class (modulo the degree of the rotation) of the chirality of the nodal point with the eigenvalues of the rotation operator. This result enhances and generalizes the description presented in reference \cite{Multi-Weyl-Topological}.

%We report a Pd3N material, where these band crossings are realized near the Fermi energy.

\section{Chiralities of band crossings due to screw rotation symmetries}
\subsection{Screw rotation symmetries}

We will be interested in screw rotation symmetries in nonsymmorphic materials; these symmetries are the composition
of a rotation around an axis and a translation along the axis of rotation. Denote by $Q$ the geometrical
operator acting on the space coordinates, $n$ the degree of the rotation and $\mathbf{a}$ the Bravais vector
around which $Q$ rotates. The screw rotation equation is thus $Q^n = p\mathbf{a}$
implying that $Q^n$ translates $p$-times the vector $\mathbf{a}$. The operator $Q$ acting on momentum coordinates
rotates around the reciprocal vector $\mathbf{b}$ of $\mathbf{a}$ and $Q^n( \mathbf{k})= \mathbf{k}$.

Denote by $\widehat{Q}$ the operator acting on the quantum states that lifts $Q$. By Bloch's theorem
we have 
\begin{align}
\widehat{Q}^n = - e^{-i p  \mathbf{k} \cdot \mathbf{a} }
\end{align}
where the negative sign is due to the spin-orbit coupling and $e^{-i p \mathbf{a} \cdot \mathbf{k}}$ is the
operator in reciprocal coordinates associated to the translation by $p\mathbf{a}$.

Consider a high symmetry line in reciprocal space starting in $\mathbf{k}^0$ and ending
in $\mathbf{k}^1$ which is fixed by $Q$, i.e. $Q \left(\mathbf{k}\right) = \mathbf{k}$ for all $\mathbf{k}$ 
on the line defined by $\mathbf{k}^0$ and
$\mathbf{k}^1$ . Assume furthermore that the end points $\mathbf{k}^0$ and
$\mathbf{k}^1$ are high symmetry points of the material such that there are no other high symmetry points in between
and
\begin{align}
e^{-i p  \mathbf{k}^0 \cdot \mathbf{a} } = \pm 1, \ \ \ \ \ e^{-i p  \mathbf{k}^1 \cdot \mathbf{a} } = \pm 1.
\end{align}

Since $\widehat{Q}$ commutes with the Hamiltonian, we may write the eigenfunction equation
\begin{align} \label{Eigenfunction of Q}
\widehat{Q} | \psi_l(\mathbf{k}) \rangle = e^{i\frac{\pi}{n}l} e^{-i \frac{p}{n}  \mathbf{k} \cdot \mathbf{a} }| \psi_l(\mathbf{k}) \rangle
\end{align}
for $|\psi_l(\mathbf{k}) \rangle$ a Bloch state for the Hamiltonian and $l = 0,1,...,2n-1$. Depending on whether $e^{-i p  \mathbf{k}^0 \cdot \mathbf{a} }$ is $1$ or $-1$, the number $l$ is odd or even
respectively. 

Screw rotations induce topologically protected band crossings along the high symmetry lines of the Brillouin zone fixed by the screw rotation
\cite{Topological-crossings-trigonal, Topological-crossings-hexagonal,Gonzalez-Tuiran-Uribe, n-hourglass}. The nodal points that appear at the intersection of the band crossings may
have different chiralities. In what follows we will study these chiralities. 

\subsection{Topologically enforced energy crossings}
\begin{figure}
	\includegraphics[width=8.8cm]{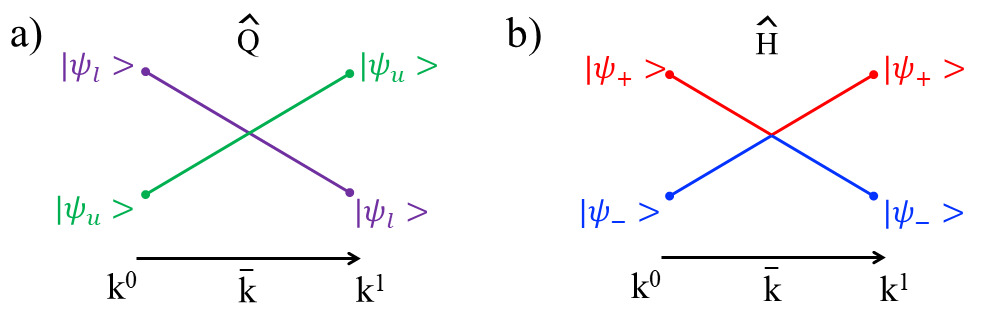}
	\caption{a) Schematic illustration of the physical effect of a) the screw rotation operator $\widehat{Q}$ and b) the hamiltonian operator  $\widehat{H}$ on the wavefunctions. On a) the colors denote the two different eigenvalues of
the screw rotation operator $\widehat{Q}$ parametrized by $u$ and $l$ once the phase $e^{-i \frac{p}{n}  \mathbf{k} \cdot \mathbf{a}}$ is removed. On b) the colors denote
the different energy bands parametrized by $+$ and $-$. Note that the eigenvalues of $\widehat{Q}$ change on the bands once they cross the critical value. Here $\overline{\mathbf{k}}$ is the coordinate where the bands intersect.} \label{Band energy crossing} 
\end{figure}

We are interested in topologically enforced crossings along the path defined by $\mathbf{k}^0$ and $\mathbf{k}^1$
which have the form presented in Fig. \ref{Band energy crossing}. Here $\psi_l$ and $\psi_u$ are simultaneous
eigenfunctions of the Hamiltonian and $\widehat{Q}$ satisfying equation \eqref{Eigenfunction of Q}.

The Hamiltonian permits us to enumerate the eigenfunctions starting with the one of less energy.
In the energy crossing presented in Fig. \ref{Band energy crossing} the Hamiltonian separates
the lower energy eigenfunction $\psi_-$ from the upper energy eigenfunction $\psi_+$, except at the energy crossing $\overline{\mathbf{k}}$
where the Hamiltonian is degenerate. Whenever the eigenvalues of $\widehat{Q}$ at $\overline{\mathbf{k}}$ agree, i.e. when
$u=l$, 
 Von Neumann and Wigner's non-crossing rule \cite{Neumann1929}  tells us that the energy bands will separate avoiding the crossing. Whenever the eigenvalues of 
$\widehat{Q}$ at $\overline{\mathbf{k}}$ differ, we thus have an enforced topological nodal point. The chirality
of such nodal point is intimately related to the eigenvalues of $\psi_l$ and $\psi_u$ along the path. We claim the following result.
 %CITA DE [J. von Neumann and E. Wigner, Z. Physik 30, 467(1929)] 

\vspace{0.5cm}

\begin{center}
{\it The chirality of the nodal point at $\overline{\mathbf{k}}$ is congruent with $\frac{l-u}{2}$ modulo $n$}
\end{center}
\vspace{0.5cm}

Recall that the chirality of the nodal point localized in $\overline{\mathbf{k}}$ is calculated as the integral
\begin{align}
\mathrm{chirality} \ \mathrm{at} \  \overline{\mathbf{k}} =\frac{1}{2\pi} \int_{S^2_{ \overline{\mathbf{k}}}} d \mathbf{S} \cdot \Omega
\end{align}
where $\Omega = \nabla \times \mathbf{A}$ is the curvature of the Berry connection $\mathbf{A} = i \langle \psi_+ | \nabla_{\mathbf{k}} | \psi_+ \rangle$ with $\psi_+$ representing the eigenfunction of the Hamiltonian with upper energy eigenvalue, and
$S^2_{ \overline{\mathbf{k}}}$ is a small 2-dimensional sphere centered at $\overline{\mathbf{k}}$ which does not 
include any other nodal point.

The sphere $S^2_{ \overline{\mathbf{k}}}$ is naturally endowed with an action of the cyclic group generated by $Q$. This action is simply a rotation around the axis
defined by the points $\mathbf{k}^0$ and $ \mathbf{k}^1$ and therefore there are only two fixed points of this action on the sphere $S^2_{ \overline{\mathbf{k}}}$
which we will denote $\overline{\mathbf{k}}^0$ and $ \overline{\mathbf{k}}^1$ respectively. For simplicity the reader may imagine
that the two points are the south and north pole respectively of a sphere and the action is rotation around the $z$-axis. For the topological 
analysis both approaches are equivalent.

Since over the sphere $S^2_{ \overline{\mathbf{k}}}$ the upper and lower eigenvalues of the Hamiltonian are gapped, we will restrict attention to the upper eigenfunction $\psi_+$ of the Hamiltonian constrained to the sphere.

Now, restricted to the sphere we may carry out a change of phase for the operator $\widehat{Q}$. Let's define the operator
\begin{align}
\widetilde{Q} =  e^{i \frac{p}{n}  \mathbf{k} \cdot \mathbf{a} } \widehat{Q}
\end{align}
and notice that with this change of phase we obtain 
\begin{align} \label{Eigenfunction of Q tilde}
\widetilde{Q} | \psi_s(\mathbf{k}) \rangle = e^{i\frac{\pi}{n}s} | \psi_s(\mathbf{k}) \rangle
\end{align}
for $\mathbf{k}$  either $\overline{\mathbf{k}}^0$ or  $\overline{\mathbf{k}}^1$ and $s = 0,1,...,2n-1$; therefore we obtain $\widetilde{Q}^n=e^{i \pi l}= \pm 1$. With this change of phase, the operator 
$\widetilde{Q}$ acts on $\psi_\pm( \overline{\mathbf{k}}^0)$ and $\psi_\pm( \overline{\mathbf{k}}^1)$ as follows:
\begin{align}
\widetilde{Q} | \psi_+( \overline{\mathbf{k}}^0) \rangle = e^{i\frac{\pi}{n}l} | \psi_+( \overline{\mathbf{k}}^0) \rangle,
 \label{Q tilde eigenvalue +0} \\
\widetilde{Q} | \psi_+( \overline{\mathbf{k}}^1) \rangle = e^{i\frac{\pi}{n}u} | \psi_+( \overline{\mathbf{k}}^1) \rangle,
 \label{Q tilde eigenvalue +1}\\ 
\widetilde{Q} | \psi_-( \overline{\mathbf{k}}^0) \rangle = e^{i\frac{\pi}{n}u} | \psi_-( \overline{\mathbf{k}}^0) \rangle,
 \label{Q tilde eigenvalue -0}\\
\widetilde{Q} | \psi_-( \overline{\mathbf{k}}^1) \rangle = e^{i\frac{\pi}{n}l} | \psi_-( \overline{\mathbf{k}}^1) \rangle.
 \label{Q tilde eigenvalue -1}
\end{align}

The eigenfunction $\psi_+$ restricted to the sphere $S^2_{ \overline{\mathbf{k}}}$ defines a section of a complex line bundle where $Q$ acts 
on the underlying sphere and $\widetilde{Q}$ acts on the section. A line bundle of this kind, with the action of
$\widetilde{Q}$ described in equations \eqref{Q tilde eigenvalue +0} and \eqref{Q tilde eigenvalue +1}, must have a
Chern number (which is the same as the chirality) which is congruent to $\frac{l-u}{2}$ modulo $n$.

To prove these results, we need to recall a result on the classification of equivariant line bundles and 
we need to construct the explicit Hamiltonian that locally models the behavior described above.

\subsection{Classification of equivariant line bundles}

The isomorphism classes of complex line bundles over a compact and oriented manifold $M$ form a group.
Line bundles may be tensored thus producing another line bundle, and line bundles have a dual line bundle which up to
isomorphism is its inverse with respect to the tensor product. The group of isomorphism classes of line bundle is called
the Picard group of $M$ and it is denoted by $Pic(M)$. A line bundle $L$ over  $M$ has a characteristic class $c_1(L) \in H^2(M, \mathbb{Z})$, called the 
Chern class of the line bundle, thus defining a second cohomology class. The map that assigns a line bundle  $L$ its Chern class $c_1(L)$
is an isomorphism of groups:
\begin{align}
Pic(M) \stackrel{\cong}{\to} H^2(M, \mathbb{Z}), \ \ [L] \mapsto c_1(L).
\end{align}
In the case of $M=S^2$ we have that $Pic(S^2) \cong \mathbb{Z}$ since $H^2(S^2, \mathbb{Z})=\mathbb{Z}$ and the Chern number
of the canonical line bundle $\gamma$ over $\mathbb{C}P^1 \cong S^2 $ is $1$.

Whenever $M$ is endowed with an action of a compact Lie group $G$ one can consider $G$-equivariant complex line bundles over $M$.
That is a complex line bundle $L$ over $M$ with an action of $G$ such that the action of $G$ is complex linear on the fibers of $L$.
The isomorphism classes of $G$-equivariant complex line bundles over $M$ becomes a group, the tensor product and the dual line bundle
define the product and the inverse respectively, and this group is denoted $Pic_G(M)$, the Picard group of $G$-equivariant complex lines bundles over $M$.  It is known \cite[Prop. 6.3]{Atiyah-Segal_Twisted_K-theory} that this group is isomorphic to $H^2(M_G, \mathbb{Z})$,
the second cohomology group of the homotopy quotient $M_G:=M \times_G EG$:
\begin{align}
Pic_G(M) \stackrel{\cong}{\to} H^2(M_G, \mathbb{Z}), \ \ L \mapsto c_1(L_G).
\end{align}

The case of interest is $M=S^2$ and $G$ the group generated by $\widetilde{Q}$, i.e $= \mathbb{Z}/n$ 
whenever $\widetilde{Q}^n=1$ and $G=\mathbb{Z}/2n$ whenever $\widetilde{Q}^n=-1$. In both cases the action
of $\widetilde{Q}$ on $S^2$ is given by the rotation action defined by $Q$. Let us assume that $\widetilde{Q}^n=1$ since
the other case is equivalent.

Applying the Serre spectral sequence \cite{Spanier} to the fiber bundle $S^2 \to S^2 \times_G EG \to BG$,  we obtain the short exact sequence
\begin{align}
0 \to H^2(BG,\mathbb{Z})  \to H^2(S^2_G, \mathbb{Z}) \to H^2(S^2,\mathbb{Z})^G \to 0 \label{short exact sequence Pic_G(S2)}
\end{align}
where $H^2(BG,\mathbb{Z}) \cong \mathrm{Hom}(G, U(1)) \cong \mathbb{Z}/n$ is the group of
1-dimensional complex representations of $G$, and  $H^2(S^2,\mathbb{Z})^G  =  H^2(S^2,\mathbb{Z})\cong \mathbb{Z}$
since the rotation action is trivial in cohomology. The short exact sequence in equation \eqref{short exact sequence  Pic_G(S2)}
splits, and therefore we have that
\begin{align}
Pic_G(S^2) \cong \mathbb{Z}/n \oplus \mathbb{Z}.
\end{align}
An explicit choice of generators is defined as follows. Denote by $C:G \to U(1)$ the 1-dimensional complex representation
defined by $C(Q^s)=e^{i\frac{2 \pi s}{n}}$. Its powers $C^j$ define the representations $C^j(Q^s)=e^{i\frac{2 \pi s j}{n}}$
with $C^n$ the trivial representation. Hence $C$ generates the group $\mathrm{Hom}(G, U(1))$.

The 2-dimensional sphere is diffeomorphic to the complex projective space $\mathbb{C}P^1$. The complex
manifold $\mathbb{C}P^1$ may be covered with
 two open spaces $U_0$ and $U_1$ with
$U_0 \cong \mathbb{C} \times\{0\}$ and $U_1 \cong \mathbb{C} \times \{1\}$ such that their intersection $U_0 \cap U_1$
induce the gluing conditions $(z,0) \sim (z^{-1},1)$. The action of $G$ on $\mathbb{C} P^1$ is given by the equation
\begin{align}
Q(z,0) = (e^{i \frac{2\pi}{n}}z, 0) \sim (e^{-i \frac{2\pi}{n}}z^{-1}, 1)=Q(z^{-1}, 1).
\end{align}
The $m$-th power $\gamma^{\otimes m}$ of the canonical line bundle $\gamma$ 
trivializes locally $\gamma^{\otimes m}|_{U_s} \cong \mathbb{C} \times \mathbb{C} \times\{s\}$, $s=0,1$ and the gluing functions
become:
\begin{align}
(\lambda, z, 0) \sim (\lambda z^m,z^{-1},1).
\end{align}
The Chern number of $\gamma^{\otimes m}$ is the winding number of the clutching function $z \mapsto z^m$ which is $m$.

In principle one can take any action of $G$ on $\gamma^{\otimes m}|_{U_0}$, but this action determines the $G$ action on
$\gamma^{\otimes m}|_{U_1}$ and therefore in all $\gamma^{\otimes m}$.
Therefore we may define the $G$ action on $\gamma^{\otimes m}$ by the equations:
\begin{align}
Q(\lambda, z, 0)= (\lambda, e^{i \frac{2\pi}{n}}z,0) \sim & \\
 ((e^{i \frac{2\pi m}{n}}\lambda z^m, e^{-i \frac{2\pi}{n}}z^{-1}, 1) &= Q(\lambda z^m,z^{-1},1)
\end{align}
and note that any other action of $G$ on  $\gamma^{\otimes m}$ is simply obtained by the tensor product $C^j \otimes \gamma^{\otimes m}$.
The explicit choice of generators of $Pic_G(S^2)$ are then $C$ and $\gamma$ and any element  is of the form $C^j \otimes \gamma^{\otimes m}$ with $(j,m) \in \mathbb{Z}/n \times \mathbb{Z}$.

 Let $(0,1)$ and $(0,0)$ be the fixed points
of the $G$ action on $\mathbb{C}P^1$ (the south and the north pole respectively) and restrict the action on
the elements of $Pic_G(S^2)$ to these two points. In both points we obtain representations of $G$. The restriction map
becomes
\begin{align}
Pic_G(S^2) \to \mathbb{Z}/n \times  \mathbb{Z}/n, \ \ C^{j} \otimes \gamma^{\otimes m}\mapsto (C^{j+m}, C^{j}).
\end{align}
Since $j+m$ and $j$ are to be considered as integers modulo $n$, we see that the
 Chern number of $C^j \otimes \gamma^{\otimes m}$, which is $m$, is congruent with $(j+m)-j$ modulo $n$. Hence we have shown
the following result:

\noindent {\bf{Theorem.}} {\it Let $L$ be a $G$-equivariant complex line bundle over $S^2$ and let $e^{i \frac{2\pi \nu}{2n}}$ and $e^{i \frac{2\pi \mu}{2n}}$
be the eigenvalues of $Q$ on the south and north pole respectively. Then the Chern number of $L$ is congruent with $\frac{\nu-\mu}{2}$ modulo $n$.}

We have therefore settled that the congruence class of the chirality of the nodal point $\overline{\mathbf{k}}$ presented above is 
congruent to $\frac{l-u}{2}$ modulo $n$.

\subsection{$G$-equivariant Hamiltonian}

Let us suppose that $\overline{\mathbf{k}}$ is centered at the origin and $Q$ is a rotation of $\frac{2\pi}{n}$ radians around the $k_z$-axis. Consider the Hamiltonian 
\begin{align} \label{Hamiltonian in k chirality m}
H(k_x,k_y,k_z)=  \left(
\begin{matrix}
k_z & (k_x-ik_y)^m \\
 (k_x+ik_y)^m & -k_z
\end{matrix}
\right)
\end{align}
and write it in cylindrical coordinates $r e^{i \phi} = k_x+ik_y$. Define the change of variables $\overline{r}=r^m$ and take
$\overline{\rho}^2=z^2+\overline{r}^2$ together with $\overline{\rho}\cos(\overline{\theta})=z$ and 
$\overline{\rho}\sin(\overline{\theta})=\overline{r}$. The Hamiltonian in the coordinates $(\overline{\rho}, \overline{\theta}, \phi)$
becomes 
\begin{align} \label{Hamiltonian overline spherical}
H(\overline{\rho}, \overline{\theta}, \phi)= \overline{\rho} \left(
\begin{matrix}
\cos(\overline{\theta}) & \sin(\overline{\theta})e^{-i m \phi} \\
\sin(\overline{\theta})e^{i m \phi} & -\cos(\overline{\theta})
\end{matrix}
\right).
\end{align}
The eigenfunctions of the transformed Hamiltonian are
\begin{align} \label{psi + psi -}
\psi_+= \left( \begin{matrix} \cos(\overline{\theta}/2)e^{-im\phi} \\ \sin(\overline{\theta}/2) \end{matrix} \right) 
\ \ \
\psi_-= \left( \begin{matrix} \sin(\overline{\theta}/2)e^{-im\phi} \\ -\cos(\overline{\theta}/2) \end{matrix} \right) 
\end{align}
whose eigenvalue equations are
\begin{align}
H | \psi_{\pm} \rangle = \pm \overline{\rho} | \psi_{\pm} \rangle =  \pm \sqrt{k_z^2 + (k_x^2+k_y^2)^{m}} |  \psi_{\pm} \rangle.
\end{align}

The action of the rotation group on the Hamiltonian must satisfy the equation:
\begin{align}
H(\mathbf{k}) = \widehat{Q}^{-1} H(Q(\mathbf{k})) \widehat{Q},
\end{align} which in the coordinates $(\overline{\rho}, \overline{\theta}, \phi)$ becomes the equation:
\begin{align}
H(\overline{\rho}, \overline{\theta}, \phi) = \left( \begin{matrix} e^{i \frac{2 \pi}{n}m} & 0 \\ 0 & 1  \end{matrix} \right)
H\left(\overline{\rho}, \overline{\theta}, \phi + \frac{2 \pi}{n}\right) 
\left( \begin{matrix} e^{-i \frac{2 \pi}{n}m} & 0 \\ 0 & 1  \end{matrix} \right),
\end{align}
thus implying:
\begin{align}
\widehat{Q} | \psi_{\pm}(\overline{\rho}, \overline{\theta}, \phi) \rangle = | \psi_{\pm}(Q(\overline{\rho}, \overline{\theta}, \phi)) \rangle,
\end{align}
which at the level of the eigenfunction $\psi_+$ becomes: 
\begin{align}
\widehat{Q}  \left( \begin{matrix} \cos(\overline{\theta}/2)e^{-im\phi} \\ \sin(\overline{\theta}/2) \end{matrix} \right)  =  \left( \begin{matrix} \cos(\overline{\theta}/2)e^{-im\phi}e^{-i \frac{2 \pi}{n}m} \\ \sin(\overline{\theta}/2) \end{matrix} \right).
\end{align}
Focusing our attention to the upper band $\psi_+$, we see that on the north pole, i.e. $\overline{\theta}=0$, the action of $\widehat{Q}$
is given by multiplication of $e^{-i \frac{2 \pi}{n}m}$, while at the south pole, i.e. $\overline{\theta}=\pi$, the action is trivial.

Let us now calculate the chirality associated to the eigenfunction $\psi_+$ of \eqref{psi + psi -} in spherical coordinates.
 The Berry connection is defined as $\mathbf{A} = i \langle \psi_+ | \nabla | \psi_+ \rangle$ with 
$\nabla = ( \partial_{\overline{\rho}},(1/ \overline{\rho}) \partial_{\overline{\theta}}, 1/(\overline{\rho} \sin{\overline{\theta}}) \partial_\phi)$ and therefore the three components of the connection become
\begin{align}
(A_{\overline{\rho}}, A_{\overline{\theta}}, A_\phi)= \left( 0,0, -\frac{m \cos^2(\overline{\theta}/2)}{\overline{\rho} \sin(\overline{\theta})}\right).
\end{align}
The Berry curvature on the spherical coordinates $(\overline{\rho}, \overline{\theta}, \phi)$ becomes: 
\begin{align}
\Omega =\nabla \times \mathbf{A} = \frac{m}{2\overline{\rho}^2} \widehat{\overline{\rho}},
\end{align}
and therefore the chirality of $\psi_+$ is:
\begin{align}
m = \frac{1}{2 \pi}\int_{S^2} d \mathbf{S} \cdot \Omega.
\end{align}

We have then that the chirality of $\psi_+$ on the cylindrical coordinates $(z, \overline{r}, \phi)$ is $m$. Now, since
the degree of the map $(z,r, \phi) \to (z, \overline{r}, \phi)$ which sends $r^m \to \overline{r}$ is $1$,
we can conclude that the chirality of $\psi_+$ on the Hamiltonian defined in \eqref{Hamiltonian in k chirality m} is also $m$. 

It is worth pointing out that the Hamiltonian of \eqref{Hamiltonian in k chirality m} appears also in equation (43) of \cite{Z2Pack} where the chiralities for $m=1,2,3$ have been calculated by measuring
the evolution of the polarization thus agreeing with the theoretical prediction.

We see also here that on the band defined by $\psi_+$, whose chirality is $m$, the action of $\widehat{Q}$ on 
$\psi_+$ whenever $\overline{\theta}=0$ is given by multiplication of $e^{-i \frac{2 \pi}{n}m}$, and whenever $\overline{\theta}=\pi$
the action is trivial. Note moreover that the action of $\widehat{Q}$ on $\psi_+$ could be tensored with any representation $C^j$, thus making
the action 
\begin{align}
\widehat{Q} = \left( \begin{matrix} e^{-i \frac{2 \pi}{n}(m-j)} & 0 \\ 0 & e^{i \frac{2 \pi}{n}j}  \end{matrix} \right).
\end{align}
The eigenvalues of $\widehat{Q}$ on $\psi_+$ and $\psi_-$ become
\begin{align}
\langle \psi_+ | \widehat{Q} | \psi_+ \rangle = \cos^2(\overline{\theta}/2) e^{-i \frac{2 \pi}{n}(m-j)} + \sin^2(\overline{\theta}/2) e^{i \frac{2 \pi}{n}j} \label{<+|Q|+>} \\
\langle \psi_- | \widehat{Q} | \psi_- \rangle =   \cos^2(\overline{\theta}/2) e^{i \frac{2 \pi}{n}j}
+\sin^2(\overline{\theta}/2) e^{-i \frac{2 \pi}{n}(m-j)} \label{<-|Q|->}
 \end{align}
 and therefore the eigenvalue of $\widehat{Q}$ on $\psi_+$ whenever $\overline{\theta}=0$ is $e^{i \frac{2 \pi}{n}(-m+j)}$ and  whenever $\overline{\theta}=\pi$ is $e^{i \frac{2 \pi}{n}j}$. 
We evidence again that $m$ is congruent to $j -(-m+j)$ modulo $n$.

 In order to obtain the chirality modulo $n$ from equations \eqref{<+|Q|+>} and \eqref{<-|Q|->} we see that
\begin{align}
\frac{n}{2\pi}\arg  & \left(\frac{\langle \psi_- | \widehat{Q} | \psi_- \rangle }{ \langle \psi_+ |   \widehat{Q} | \psi_+     \rangle}  \right)    \equiv _n \left\{ \begin{matrix} m & \mbox{whenever} & \overline{\theta}=0\\
 -m & \mbox{whenever} & \overline{\theta}=\pi.\\ \end{matrix}  \right.  \label{equation chirality} 
\end{align}
Using the notation of equation \eqref{Eigenfunction of Q tilde}, which refers to Fig. \ref{Band energy crossing}a),
 we see that $\frac{l}{2}=j$ and $\frac{u}{2}=-m+j$. Moreover we have that 
\begin{align}
\frac{n}{2\pi} \arg & \left( \frac{\langle   \psi_-(\mathbf{k})| \widehat{Q} | \psi_- (\mathbf{k})\rangle}{\langle \psi_+(\mathbf{k}) |   \widehat{Q} | \psi_+(\mathbf{k}     \rangle}  \right)\label{formula chirality between bands}  \\
 &  = \left\{ \begin{matrix} \frac{l-u}{2} \equiv_n m & \mbox{for} & \mathbf{k} \ \mbox{between} \  \overline{\mathbf{k}} \ \mbox{and} \ \mathbf{k}^1,\\
-\frac{l-u}{2} \equiv_n -m & \mbox{for} & \mathbf{k} \ \mbox{between} \  \ \mathbf{k}^0 \ \mbox{and}    \ \overline{\mathbf{k}}. \end{matrix}  \right.    \nonumber
\end{align}

Therefore, if equation \eqref{formula chirality between bands} is plotted as a function of the $\mathbf{k}$ coordinate, a step-function type with the chirality $m$ of the nodal point should be expected. This behavior can be observed in Fig. \ref{figure example chi} where equation \eqref{formula chirality between bands} is plotted as a function of $k_z$ for two successive bands of a P$6_1$22 material ($n$=6). In this case, a chirality of $-2$ is obtained at the right of the nodal point located at $\overline{\mathbf{k}}$. 

It is important to emphasize that the expression \eqref{formula chirality between bands} can be used to determine the chiralities of band crossings in screw symmetric materials by using only the eigenvalues of the screw operator $\widehat{Q}$ along the high symmetry
 lines that the operator leaves
invariant.

\section{Chiralities along hourglass and accordion like band diagrams}

In the presence of screw rotation symmetries of the type P2$_1$, P3$_p$, P4$_p$ and P6$_p$, the formation
of hourglass and accordion like shapes on the energy bands along the screw-invariant lines has been described in \cite{Topological-crossings-trigonal,Topological-crossings-hexagonal,Gonzalez-Tuiran-Uribe}. These combinatorial formations predict topological energy band crossings whose
chirality can be calculated with the description defined previously.
  
For simplicity let us assume that our geometrical operator $Q$ does a screw rotation along the $z$-axis. Therefore
we will take $\mathbf{k}^0= \Gamma$ and  $\mathbf{k}^1= A$ with the eigenfunction equation  \eqref{Eigenfunction of Q}
thus becoming
\begin{align} \label{Eigenfunction Q on kz}
\widehat{Q} | \psi_l(\mathbf{k}) \rangle = e^{i\frac{\pi}{n}l} e^{-i k_z \frac{p}{n} }| \psi_l(\mathbf{k}) \rangle
\end{align}
with $\widehat{Q}^n=-e^{-i k_z p}$ for $n=2,3,4,6$. Kramer's degeneracy rule on $\Gamma$ and $A$ forces the eigenvalues of $\widehat{Q}$ to 
appear in conjugate pairs, and the nonsymmorphicity of $\widehat{Q}$ is the key ingredient for the formation of hourglass
and accordion shape energy band diagrams.

In the following combinatorial band diagrams we have written the eigenvalues of the operator $\widehat{Q}$ on
$\Gamma$ and $A$ and the chiralities of the energy crossings have to be chosen so that their absolute value is minimum. The chiralities will
be printed in red above the intersection of the bands. The numbers $n=2,3,4,6$ and $1\leq p <n$ encode the
information of the screw rotation. We remark here that the band diagrams for the symmetry groups
P6$_p$ are equivalent to the ones of P6$_p$22 (see Appendix in \cite{Gonzalez-Tuiran-Uribe}) and
the same argument implies that the diagrams for P4$_p$ are equivalent to the ones of P4$_p$22.
 For $n=2$ the chirality  is $\pm 1$:

\begin{equation} \label{chirality P21}
\xymatrix@1@R=0.12cm@C=0.5cm{
\Gamma & p=1& A\\
{{e^{i\pi 1/2} \atop e^{i\pi 3/2}}}  \bullet \ar@{-}[rr] \ar@{-}[rrd]^(.5){\textcolor{red}{\pm 1}} &&\bullet {e^{0} \atop e^{0}}\\
{e^{i\pi 1/2} \atop e^{i\pi 3/2}} \bullet \ar@{-}[rru] \ar@{-}[rr] && \bullet {e^{i\pi 2/2} \atop e^{i\pi 2/2}}
} \end{equation}

For $n=3$ the band diagrams for $p=1$ and $p=2$ are similar. The chiralities modulo 3 are:
\begin{equation} \label{chirality P31}
\xymatrix@1@R=0.12cm@C=0.5cm{
\Gamma & p=1& A\\
{{e^{i\pi 1/3} \atop e^{i\pi 5/3}}}  \bullet \ar@{-}[rr] \ar@{-}[rrd]^(.5){\textcolor{red}{- 1}} &&\bullet {e^{0} \atop e^{0}}\\
{e^{i\pi 1/3} \atop e^{i\pi 5/3}} \bullet \ar@{-}[rru] \ar@{-}[rrd]^(.5){\textcolor{red}{+ 1}} && \bullet {e^{i\pi 4/3} \atop e^{i\pi 2/3}}\\
{e^{i\pi 3/3} \atop e^{i\pi 3/3}} \bullet \ar@{-}[rru] \ar@{-}[rr] && \bullet {e^{i\pi 4/3} \atop e^{i\pi 2/3}}
} \end{equation}

For $n=4$ the band diagrams for $p=1$ and $p=3$ are similar. The chiralities modulo 4 are:
\begin{equation} \label{chiralities P41 P42}
\xymatrix@1@R=0.06cm@C=0.3cm{
\Gamma &p=1& A & \Gamma & p=2& A\\
{e^{i\pi/4} \atop e^{i\pi 7/4}}  \bullet \ar@{-}[rr] \ar@{-}[rrd]^(.5){\textcolor{red}{- 1}} &&\bullet {e^0 \atop e^0 }
    &  {e^{i\pi/4} \atop e^{i\pi 7/4}}  \bullet \ar@{-}[rr] \ar@{-}[rrd]^(.5){\textcolor{red}{\pm 2}} &&\bullet {e^{i\pi 7/4} \atop e^{i\pi 1/4}}\\
{e^{i\pi /4} \atop e^{i\pi 7/4}} \bullet \ar@{-}[rrd]^(.5){\textcolor{red}{\pm 2}} \ar@{-}[rru] && \bullet {e^{i\pi 6/4} \atop e^{i\pi 2/4}}
   &  {e^{i\pi 3/4} \atop e^{i\pi 5/4}} \bullet \ar@{-}[rr] \ar@{-}[rru] && \bullet {e^{i\pi 5/4} \atop e^{i\pi 3/4}}\\
{e^{i\pi 3/4} \atop e^{i\pi 5/4}} \bullet \ar@{-}[rrd]^(.5){\textcolor{red}{+ 1}} \ar@{-}[rru]&& \bullet {e^{i\pi 6/4} \atop e^{i\pi 2/4}}
   & && \\
{e^{i\pi 3/4} \atop e^{i\pi 5/4}} \bullet \ar@{-}[rr] \ar@{-}[rru] && \bullet {e^{i\pi 4/4} \atop e^{i\pi 4/4}}
    & && 
} \end{equation}

For $n=6$ the band diagrams for $p=1$ and $p=5$ and for $p=2$ and $p=4$ are similar. The chiralities modulo 6 are:

\begin{equation} \label{chiralities P61 P62 P63}
\xymatrix@1@R=0.06cm@C=0.3cm{
\Gamma &p=1& A & \Gamma &p=2& A\\
{e^{i\pi/6} \atop e^{i\pi 11/6}}  \bullet \ar@{-}[rr] \ar@{-}[rrd]^(.5){\textcolor{red}{- 1}} &&\bullet {e^0 \atop e^0 }
    &  {e^{i\pi/6} \atop e^{i\pi 11/6}}  \bullet \ar@{-}[rr] \ar@{-}[rrd]^(.5){\textcolor{red}{- 2}} &&\bullet {e^{i\pi 11/6} \atop e^{i\pi 1/6}}\\
{e^{i\pi /6} \atop e^{i\pi 11/6}} \bullet \ar@{-}[rrd]^(.5){\textcolor{red}{- 2}} \ar@{-}[rru] && \bullet {e^{i\pi 10/6} \atop e^{i\pi 2/6}}
   &  {e^{i\pi 3/6} \atop e^{i\pi 9/6}} \bullet \ar@{-}[rrd]^(.5){\textcolor{red}{+ 2}} \ar@{-}[rru] && \bullet {e^{i\pi 9/6} \atop e^{i\pi 3/6}}\\
{e^{i\pi 3/6} \atop e^{i\pi 9/6}} \bullet \ar@{-}[rrd]^(.5){\textcolor{red}{\pm 3}} \ar@{-}[rru]&& \bullet {e^{i\pi 10/6} \atop e^{i\pi 2/6}}
   & {e^{i\pi 5/6} \atop e^{i\pi 7/6}} \bullet \ar@{-}[rr] \ar@{-}[rru]&& \bullet {e^{i\pi 7/6} \atop e^{i\pi 5/6}}\\
{e^{i\pi 3/6} \atop e^{i\pi 9/6}} \bullet \ar@{-}[rrd]^(.5){\textcolor{red}{+ 2}} \ar@{-}[rru] && \bullet {e^{i\pi 8/6} \atop e^{i\pi 4/6}}
    & \Gamma & p=3& A \\
{e^{i\pi 5/6} \atop e^{i\pi 7/6}} \bullet \ar@{-}[rrd]^(.5){\textcolor{red}{+ 1}} \ar@{-}[rru] && \bullet {e^{i\pi 8/6} \atop e^{i\pi 4/6}}
   & {e^{i\pi /6} \atop e^{i\pi 11/6}} \bullet \ar@{-}[rrd]^(.5){\textcolor{red}{\pm 3}} \ar@{-}[rr] && \bullet {e^{i\pi 10/6} \atop e^{i\pi 2/6}} \\
{e^{i\pi 5/6} \atop e^{i\pi 7/6}} \bullet \ar@{-}[rr] \ar@{-}[rru] && \bullet {e^{i\pi 6/6} \atop e^{i\pi 6/6}}
   & {e^{i\pi 5/6} \atop e^{i\pi 7/6}} \bullet \ar@{-}[rr] \ar@{-}[rru] && \bullet {e^{i\pi 8/6} \atop e^{i\pi 4/6}}\\
&& &{e^{i\pi 3/6} \atop e^{i\pi 9/6}} \bullet \ar@{-}[rrd]^(.5){\textcolor{red}{\pm 3}} \ar@{-}[rr] && \bullet {e^{0} \atop e^{0}} \\
&&&  {e^{i\pi 3/6} \atop e^{i\pi 9/6}} \bullet \ar@{-}[rr] \ar@{-}[rru] && \bullet {e^{i\pi 6/6} \atop e^{i\pi 6/6}}
} \end{equation}

The band diagrams for $n=6$ and $p=3$ might appear superposed. In that case the chiralities of all the intersections can also be calculated.
One possible way on which the band diagrams superpose is the following. The chiralities are modulo 6.

\begin{equation} \label{chiralities superposed P63}
\xymatrix@1@R=0.3cm@C=0.7cm{
\Gamma & n=6, p=3& A
\\
{e^{i\pi 7/6} \atop e^{i\pi 5/6}} \bullet \ar@{-}[rrd]^(.77){\textcolor{red}{-1}}
 \ar@{-}[rrddd]_(0.63){\textcolor{red}{\pm 3}}
 &&{\textcolor{blue}{ \bullet {e^{i\pi 6/6} \atop e^{i\pi 6/6}}}} \ar@{}[lldddd]_(.52){\textcolor{red}{+1}}
\\
{\textcolor{blue}{{e^{i\pi 9/6} \atop e^{i\pi 3/6}}\bullet}} \ar@{-}@[blue][rru]^(.25){\textcolor{red}{-2}}
 \ar@{-}@[blue][rrd]^(.47){\textcolor{red}{+1}}
 && \bullet {e^{i\pi 4/6} \atop e^{i\pi 8/6}} \ar@{}[lldd]_(.46){\textcolor{red}{-2}}  \ar@{}[llu]_(.5){\textcolor{red}{-1}}
\\
&& {\textcolor{blue}{ \bullet {e^{i\pi 0/6} \atop e^{i\pi 0/6}}}}
\\
 {e^{i\pi 11/6} \atop e^{i\pi 1/6}} \bullet \ar@{-}[rruu]^(.7){\textcolor{red}{+2}}
 \ar@{-}[rr]^(.46){\textcolor{red}{-1}}
 && \bullet {e^{i\pi 2/6} \atop e^{i\pi 10/6}}\ar@{}[ll]_(.77){\textcolor{red}{+2}}
\\
{\textcolor{blue}{{e^{i\pi 9/6} \atop e^{i\pi 3/6}}\bullet}} \ar@{-}@[blue][rruuuu]^(0.62){\textcolor{red}{\pm 3}}
 \ar@{-}@[blue][rruu]^(.82){\textcolor{red}{+1}}
 &&
 }\end{equation}

Whenever $n=6$ and $p=1$ and we consider the group P6$_1$, we can also consider the band diagram along K-H. Here the operator that fixes the line
is $\widehat{Q}^2$ and the combinatorial band diagram has been calculated in \cite{Gonzalez-Tuiran-Uribe}. The chiralities
are in red and are modulo $3$ since the order of the rotation operator $Q^2$ is $3$.
\begin{equation}  \label{chiralities P6_1 KH}
\xymatrix@1@R=0.2cm@C=1.1cm{
K &P& H\\
 {e^{i\pi 5/3} \atop e^{i\pi 1/3}} \bullet \ar@{-}[rr] \ar@{-}[rrd]_(.5){\textcolor{red}{-1}}  &&\bullet {e^{i\pi4 /3} \atop e^{i\pi 2/3}}\\
{{e^{i\pi3/3}} \atop {}} \bullet  \ar@{-}[rru]&& \bullet {e^{0} \atop e^{0}}\\
{{} \atop {e^{i\pi 3/3}}} \bullet \ar@{-}[rr]^(.5){\textcolor{red}{+1}}  && \bullet {e^{i\pi 2/3} \atop e^{i\pi 4/3}}\\
{e^{i\pi 1/3} \atop e^{i\pi 5/3}} \bullet  \ar@{-}[rru]  \ar@{-}[rruu]&& 
}
\end{equation}

The chiralities presented in red in the previous diagrams are all modulo $n$. A fair guess would be that the absolute value of the chiralities take always the minimum possible value, and therefore,
and in the appropriate cases, the congruence class determines the chirality. Unfortunately, this is not the case. In Figure \ref{AgF3-KH} we show
that the congruence class of the upper nodal point is $+1$ according to the eigenvalues of the representations, but the chirality calculated with the Wilson loop method is $-2$.
This means that the explicit chirality cannot be deduced only from the eigenvalues of the representations.
Nevertheless, the congruence class of the chirality obtained through the eigenvalues of the rotation operator puts a strong restriction on the possible values that the chirality might have, and since this calculation could be done directly from the DFT. This method provides a very simple procedure to determine the type of the nodal points along the rotation invariant lines.

\section{Materials realization}

\begin{figure}
	\includegraphics[width=8.8cm]{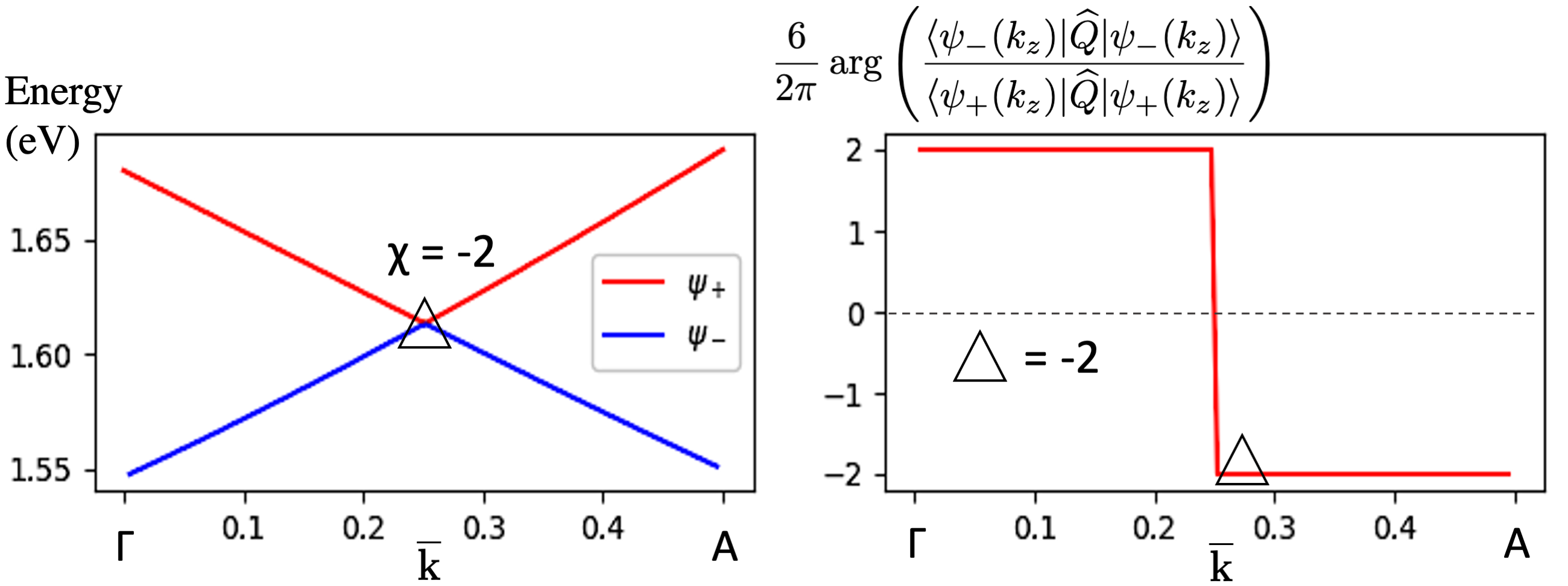}
	\caption{
		(Left) Electronic band structure for the fourth ($\psi_-$) and fifth ($\psi_+$) 
		conduction bands of AuF$_3$ (P6$_1$22) restricted to $\Gamma$-A. (Right) Graph of expression 
		\eqref{chirality succesive bands} 
		for the succesive bands $\psi_-$ and $\psi_+$ with $\widehat{Q}$ the skew symmetry operator restricted to $\Gamma$-A. The topological Weyl chirality of the nodal point $\mathbf{\overline{k}}$
		corresponds to the value of 
		\eqref{chirality succesive bands} 
		at the right of $\mathbf{\overline{k}}$. In this case
		the chirality is $-2$.
	}
	\label{figure example chi} 
\end{figure}

\begin{figure*}
	\includegraphics[width=17.8cm]{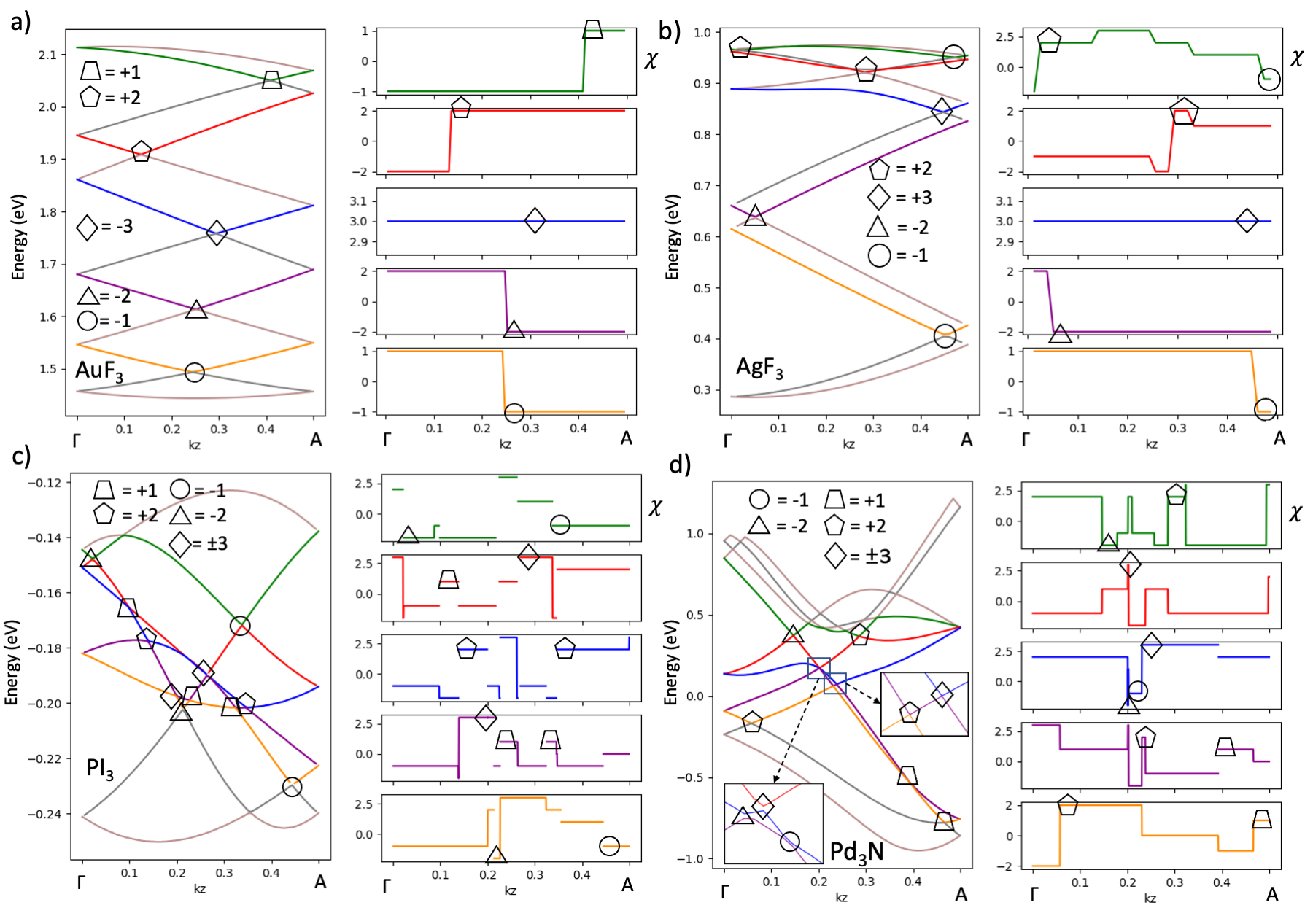}
	\caption{(Left) Electronic band structure for a) AuF$_3$ (P6$_1$22), b) AgF$_3$ (P6$_1$22), c) PI$_3$ (P6$_3$) and d) Pd$_3$N (P6$_3$22) with the topological chiralities calculated by the integration of the Berry curvature at the surface of a k-sphere around the Weyl point \cite{wanniertools}. The chiralities along $\Gamma$-A are compatible with the ones described in diagrams \eqref{chiralities P61 P62 P63} and \eqref{chirality P21} respectively. (Right) Topological Weyl chirality calculated from the eigenvalues of the screw symmetry operator \cite{irrep} along the $\Gamma$-A path using the formula of expression
		\eqref{chirality succesive bands}. The color of the chirality plot matches the energy band calculated.
	} \label{bandas} 
\end{figure*}

The chiralities previously described are compatible with the ones presented in Fig. \ref{bandas}   for the symmetry space groups P6$_1$22, P6$_3$ and P6$_3$22 of materials AgF$_3$, AuF$_3$, PI$_3$ and Pd$_3$N along 
the high symmetry line $\Gamma$-A. For the former two materials we have $n=6$ and $p=1$ and for the
latter two  $n=6$ and $p=3$. Let us elaborate.

In order to determine the chirality of the Weyl points along $\Gamma$-A we have
carried out two different procedures. On the one hand we have taken small spheres around the nodal points and we have 
calculated the Berry curvature flux through the spheres \cite{Berry-phase}. This was possible by the Wannier interpolation technique (\cite{wannier90}) allowing one to determine the Berry curvature with an efficient $k$-mesh sampling \cite{wanniertools}. The values of these chiralities are shown on the electronic band structure for the a) AuF$_3$, b) AgF$_3$, c) PI$_3$ and d) Pd$_3$N materials in the left panel of Fig. \ref{bandas}.  These results are in complete agreement with the chiralities predicted in diagrams \eqref{chiralities P61 P62 P63} from the chirality nodal point theorem.

On the other hand, we have calculated the symmetry eigenvalues of the wavefunctions at high symmetry lines in the Brillouin zone (irrep code \cite{irrep}).  This step only requires the information of the electronic wavefunction in the crystal system which can be obtained directly from density functional theory (DFT) calculations. Then, we use the eigenvalues of the screw symmetry operator $\widehat{Q}$ in order to calculate the chirality of the Weyl points using the chirality nodal point theorem through the formula \eqref{formula chirality between bands}. For each pair of consecutive bands $\psi_{-}$ and $\psi_{+}$,
we have restricted to the line $k_x=k_y=0$ and 
we have plotted the expression
\begin{align}
\frac{6}{2\pi}\arg \left(\frac{\langle \psi_{-}(k_z)| \widehat{Q} | \psi_{-} (k_z)\rangle}{\langle \psi_+(k_z) |   \widehat{Q} | \psi_+(k_z)     \rangle}\right) \label{chirality succesive bands}
\end{align}
for $0 \leq k_z \leq 0.5$. Whenever there is an energy crossing between the bands, say at $k_z=\overline{k}$, equations \eqref{equation chirality} and 
\eqref{formula chirality between bands} tell us that the chirality of the Weyl point is given by the value of the 
expression in \eqref{chirality succesive bands} at the right of $\overline{k}$.

Fig. \ref{figure example chi} shows the graph of expression \eqref{chirality succesive bands} applied to the fourth and fifth conduction bands of
AuF$_3$. The chirality of the nodal point corresponds to the value of the expression \eqref{chirality succesive bands}
on the right of $\mathbf{\overline{k}}$; in this case the chirality of the nodal point is $-2$.

The results of the chirality calculations with this procedure are presented in the right panels of Fig. \ref{bandas}. It is important
then to notice that both procedures to calculate the chiralities agree on all the nodal points we have presented. This
evidences the pertinence of the theoretical description presented above and its precise applicability. 
Moreover, this method permits  to calculate the Weyl chirality directly from DFT wavefunctions without change of basis set representation (for example, wannier representation \cite{MLWF}). In terms of computational complexity, the calculation of the chirality of nodal points using representation
theory \cite{irrep} is far simpler than the calculation of the chirality through the holonomy of the Berry connection.

From the right panels of Fig. \ref{bandas}  we have evidenced the existence of high order chiralities in the P6$_p$22 symmetry 
groups along the $\Gamma$-A high symmetry line. A particular ordered chirality sequence of $+1,+2,\pm 3,-2,-1$ was obtained for 
the accordion band structure of the AgF$_3$ and AuF$_3$ (P6$_1$22 space group).
 These results are consistent with the ones presented in diagram \eqref{chiralities P61 P62 P63} and the Weyl chiralities calculated 
from Berry curvature (left panels in Fig. \ref{bandas}  a) and b)). In addition, different P6$_1$ materials or special doping in AgF$_3$ or 
AuF$_3$ compounds, could localize the Fermi level at one specific energy position and have access to each of these particular chiralities.

\begin{figure}
	\includegraphics[width=8.8cm]{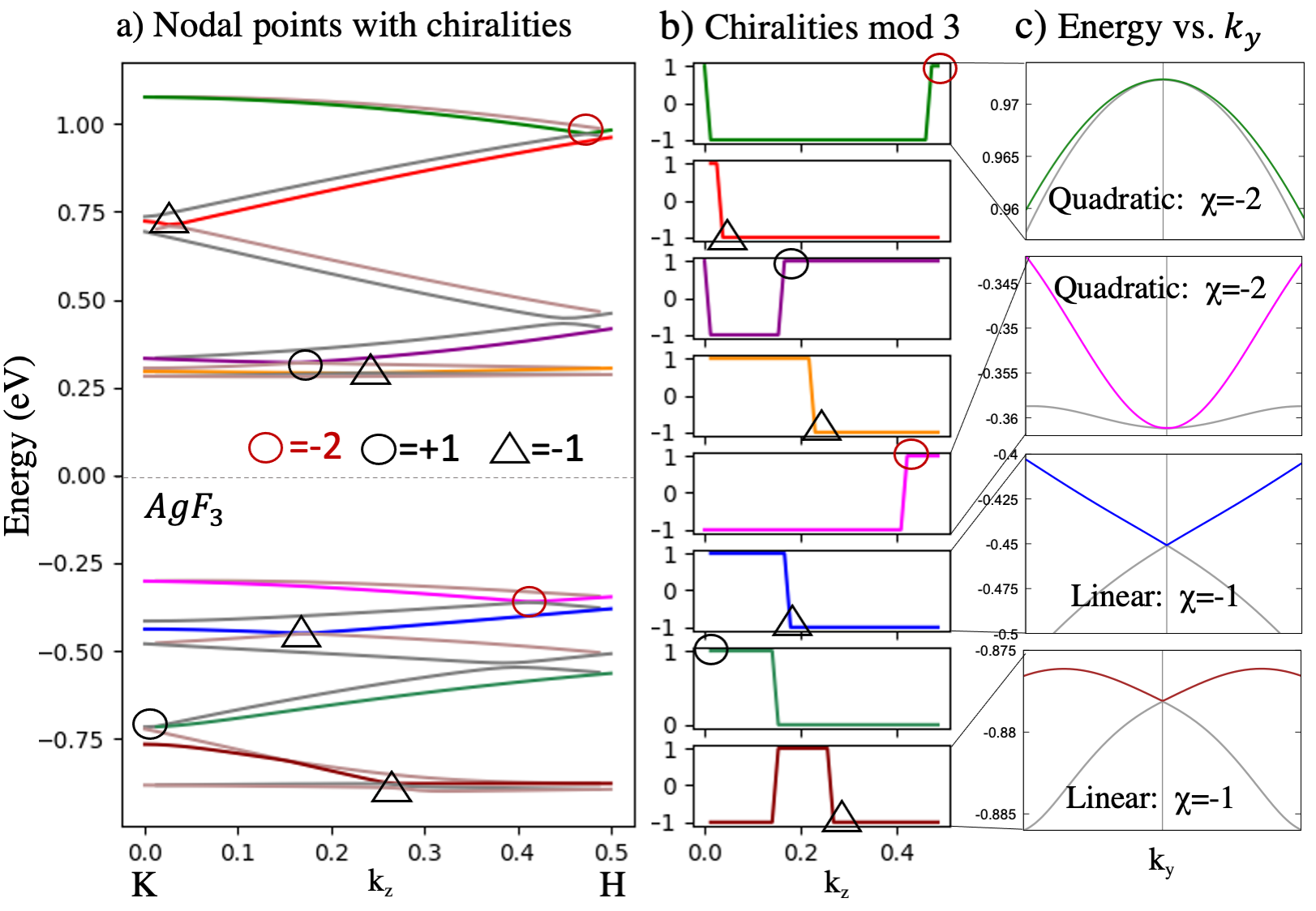}
	\caption{a) Electronic band structure for AgF$_3$ restricted to the path K-H. Topological chiralities have been calculated with the Berry curvature flux integration \cite{wanniertools}.  b) Topological Weyl chirality modulo 3 calculated from the eigenvalues of screw symmetry operator \cite{irrep} along the K-H path. The color of the chirality plot represents the energy band calculated. The patterns of the chiralities are compatible with the ones presented in diagram \eqref{chiralities P6_1 KH}. c) Energy dispersion along the $k_y$ axis for the selected four nodal points. 
It is important to notice that among the eight nodal points, six have $\pm1$ chiralities but two have $-2$ chirality.  This shows that the chirality cannot be deduced explicitly from the congruence class alone;
nevertheless the congruence class puts a strong restriction on the possible values the chirality might have.
	} \label{AgF3-KH} 
\end{figure}

In addition, triple Weyl points of $\pm$3 chirality in the $\Gamma$-A path were obtained for P6$_3$ and P6$_3$22 space groups, as it is shown in Figure \ref{bandas} c) and d) respectively.  For the Pd$_3$N case, close to the Fermi energy we found two Weyl points with +3 chirality, which could generate an exotic spin transport response as large spin Hall conductivity in this material. For the PI$_3$ case, the
configuration of Weyl points in the superposed hourglass configuration is similar to the one presented in diagram \eqref{chiralities superposed P63} and the value of the chiralities agree. The nodal points with $\pm 3$ chirality in both cases lie at the center of the hourglass configurations.

In order to study another k$_z$ high-symmetry line on the Brillouin zone, we show the results of the electronic band structure for the K-H path of the AgF$_3$ material  in Fig. \ref{AgF3-KH}. 
Here we have $n=3$ and $p=1$.  We present the last twelve conduction bands together with the first twelve valence bands and
we calculate the congruence class modulo $3$ of the chiralities of the nodal points with the procedure
outlined above. We moreover determine the explicit value of the chiralities using the Berry curvature flux integration
method \cite{wanniertools}
and we show their residue modulo $3$ agrees with the one calculated using the eigenvalues
of the rotation operator. It is important to emphasize that the congruence class
of the chirality does not determine the explicit value of the chirality. We can see in 
Fig. \ref{AgF3-KH} that among the eight nodal points presented, six have $\pm 1$ chirality, while two
have $-2$ chirality. The congruence class modulo $3$ of diagram \eqref{chiralities P6_1 KH} is the correct one, but the chirality may take values different than $\pm1$.

%---------- AL FINAL DE NUEVO
\subsection{Computational methods}
Density functional calculations (DFT) based in the Vienna ab initio package (VASP) code \cite{vasp} were performed to examine the electronic band structure and the eigenvalues of the symmetry operator \cite{irrep}. The projector augmented wave (PAW) \cite{PAW} method was adopted to treat the core-valence electron interactions. The exchange-correlation interactions were chosen within the Perdew-Berke-Ernzerhof (PBE) \cite{pbe} schemes with a cutoff energy 520 eV. The crystal structures were obtained from a relaxed system by Materials project \cite{materialsproject}. In all the calculations, spin-orbit coupling (SOC) was included self-consistently at the DFT level.  

In order to evaluate the Weyl point chirality, we have used an effective tight-binding Hamiltonian constructed in the maximally localized Wannier basis \cite{wannier90} as a post-processing step of the DFT calculations.  We have employed the Wanniertools package \cite{wanniertools} with a $k$-resolved $240^3$ $k$-mesh to obtain the Berry flux through the spheres (centered in $\bar{k}$) and calculate the chirality of Weyl points obtained in the nonsymmorphic materials PI$_3$ (SG: P6$_3$), Pd$_3$N (SG: P6$_3$22), 
AgF$_3$ (SG: P6$_1$22), AuF$_3$ (SG: P6$_1$22)..

\section{Conclusion}

Two interesting features appear in the presence of screw rotation symmetries and spin-orbit coupling whenever we look at the energy bands
on screw invariant lines. First, the distribution of the energy bands follows hourglass and accordion shape configuration schemes, thus enforcing nodal points at the band intersections; these are the topological enforced Weyl points. On the other hand, the eigenfunctions of the Hamiltonian are simultaneous eigenfunctions of the screw rotation operator, which in turn permits to extract local information of the Hamiltonian
at the band intersections. The combination of these two features permits not only determine the chirality of the topological enforced Weyl points
but also determine the chirality of any energy band intersection.

The precise form under which the screw rotation operator acts on the wavefunctions on band intersections puts strong
restrictions on the local structure of the Hamiltonian. In particular, if the eigenvalues of the screw rotation operator on the two bands
are different, the bands cannot be locally gaped, and the congruence class of the chirality is determined explicitly from the difference of the arguments of the eigenvalues. 

The theoretical and computational procedure presented in this work permits to determine the congruence class of the chiralities of nodal points on screw invariant lines through the information provided by DFT calculations. High-order band crossings (chiralities of $\pm$2 and $\pm$3) are obtained for P6$_p$
and P6$_p$22 nonsymmorphic space groups.  Of particular importance is the fact that energy crossings that are not topologically protected (as it is the case in the intersections in the hourglass and accordion shape diagrams) may be nevertheless protected by the representation theory of the screw rotation operator.

The argument presented in this work permits to evaluate the congruence class of the chiralities of Weyl points that lie on rotation and screw rotation invariant lines. The information required are the eigenvalues of the rotation or screw rotation operator on the wavefunctions.

We believe that this procedure could be computationally implemented so that the band crossings calculated through DFT would appear labeled with the corresponding chirality. We have shown in this work how this labeling would be performed. The generic implementation would be welcomed.

\section{Acknowledgments}
The first and the third authors thank the continuous support of the Alexander Von Humboldt Foundation, Germany. 
The first author gratefully acknowledges the computing time granted on the supercomputer Mogon at Johannes Gutenberg University Mainz (hpc.uni-mainz.de). The second author thanks the German Service of Academic Exchange (DAAD) for its continuous support. The third author acknowledges the support of the Max Planck Institute for Mathematics in Bonn, Germany.

\bibliographystyle{apsrev4-2}
\bibliography{topological}
%\bibliography{topological.bib}

\end{document}